\RequirePackage{fix-cm}
\documentclass[smallextended]{svjour3}       
\smartqed  
\usepackage{graphicx}
\usepackage{graphicx,subfigure}
\usepackage{array}
\usepackage{natbib}

\begin{document}

%
%
\title{{\it Herschel} SPIRE FTS Spectral Mapping Calibration\thanks{{\it Herschel} is an ESA space observatory with science instruments provided by European-led Principal Investigator consortia and with important participation from NASA.}
}


%
%
\author{Dominique Benielli \and
Edward Polehampton \and
Rosalind Hopwood \and Ana Bel\'{e}n Gri\~{n}\'{o}n Mar\'{i}n
\and Trevor Fulton  \and  Peter Imhof \and  Tanya Lim  \and Nanyao Lu \and Gibion Makiwa \and Nicola Marchili \and  David Naylor \and Locke Spencer \and Bruce Swinyard \and  Ivan Valtchanov \and Matthijs van der Wiel }


\institute{D. Benielli   \at
Aix Marseille Universit\'e, CNRS, LAM (Laboratoire d'Astrophysique de Marseille) UMR 7326, 13388, Marseille, France
           \and
E. Polehampton \at
RAL Space, Rutherford Appleton Laboratory, Didcot OX11 0QX, UK; \\
Institute for Space Imaging Science, Department of Physics \& Astronomy, University of Lethbridge, Lethbridge, AB T1K3M4, Canada
\and
R. Hopwood \at
Physics Department, Imperial College London, South Kensington Campus, SW7 2AZ, UK
\and
A. B. Gri\~{n}\'{o}n Mar\'{i}n \at
RAL Space, Rutherford Appleton Laboratory, Didcot OX11 0QX, UK
\and
T. Fulton \at
Bluesky Spectroscopy, Lethbridge, AB T1J 0N9, Canada;\\
Institute for Space Imaging Science, Department of Physics \& Astronomy, University of Lethbridge, Lethbridge, AB T1K3M4, Canada 
\and
P. Imhof \at
Bluesky Spectroscopy, Lethbridge, AB T1J 0N9, Canada;\\
Institute for Space Imaging Science, Department of Physics \& Astronomy, University of Lethbridge, Lethbridge, AB T1K3M4, Canada
\and
T. Lim \at
RAL Space, Rutherford Appleton Laboratory, Didcot OX11 0QX, UK
\and
N. Lu \at
NHSC/IPAC, 100-22 Caltech, Pasadena, CA 91125, USA
\and
G. Makiwa \at
Institute for Space Imaging Science, Department of Physics \& Astronomy, University of Lethbridge, Lethbridge, AB T1K3M4, Canada
\and
N. Marchili \at
Dipartimento di Fisica e Astronomia, Universit\`a di Padova, I-35131 Padova, Italy
\and
D. A. Naylor \at
Institute for Space Imaging Science, Department of Physics \& Astronomy, University of Lethbridge, Lethbridge, AB T1K3M4, Canada
\and
L. D. Spencer \at
Institute for Space Imaging Science, Department of Physics \& Astronomy, University of Lethbridge, Lethbridge, AB T1K3M4, Canada
\and
B. Swinyard \at
RAL Space, Rutherford Appleton Laboratory, Didcot OX11 0QX, UK
\and
I. Valtchanov \at
Herschel Science Centre, ESAC, P.O. Box 78, 28691 Villanueva de la Ca\~nada, Madrid, Spain
\and
M. H. D. van der Wiel \at
Institute for Space Imaging Science, Department of Physics \& Astronomy, University of Lethbridge, Lethbridge, AB T1K3M4, Canada
}

\date{Received: date / Accepted: date}

\maketitle

%
%
\begin{abstract}
The {\it Herschel} SPIRE Fourier transform spectrometer (FTS) performs spectral imaging in the 447--1546~GHz band. It can observe in three spatial sampling modes: sparse mode, with a single pointing on sky, or intermediate or full modes with 1 and 1/2 beam spacing, respectively. In this paper, we investigate the uncertainty and repeatability for fully sampled FTS mapping observations. The repeatability is characterised using nine observations of the Orion Bar. Metrics are derived based on the ratio of the measured intensity in each observation compared to that in the combined spectral cube from all observations. The mean relative deviation is determined to be within 2\%, and the pixel-by-pixel scatter is $\sim$7\%. The scatter increases towards the edges of the maps. The  uncertainty in the frequency scale is also studied, and the spread in the line centre velocity across the maps is found to be $\sim$15~km\,s$^{-1}$. Other causes of uncertainty are also discussed including the effect of pointing and the additive uncertainty in the continuum.

\end{abstract}

%
%
\section{Introduction}
\label{intro}
The Spectral and Photometric Imaging REceiver \citep[SPIRE;][]{griffin2010} is one of the focal plane instruments on board the ESA {\it Herschel} Space Observatory \citep{pilbratt10}. SPIRE contains an imaging photometric camera and an imaging Fourier transform spectrometer (FTS). The FTS can produce spectral images over a circular 2$^{\prime}$  field of view with two hexagonally packed bolometer arrays: SLW (447--990 GHz) and SSW (958--1546 GHz). The spectral resolution achieved in the highest resolution mode is 1.184 GHz across both bands, and the full width at half maximum (FWHM) of the beam  viewed by each bolometer ranges between  31--43 $^{{\prime}{\prime}}$ for SLW and 16--20$^{{\prime}{\prime}}$ for SSW \citep{makiwa2013}. Three spatial sampling modes are available: sparse with 2 beam spacing (2F$\lambda$), intermediate with 1 beam spacing (1F$\lambda$) and full with 1/2 beam spacing (0.5F$\lambda$). The full (Nyquist) sampling is achieved with a 16-point jiggle using the SPIRE internal beam steering mirror \citep{observersmanual}. Each jiggle position is calibrated independently, and the detector data resampled onto a regular spatial grid with pixel sizes of 9.5$^{{\prime}{\prime}}$ for SSW and 17.5$^{{\prime}{\prime}}$ for SLW. A full description of the calibration scheme is given in \citet{swinyard13}.

Intermediate and fully sampled mapping observations account for approximately 25\% of the SPIRE FTS observing time, and it is therefore important to understand the uncertainties associated with these observing modes. In this paper, we evaluate the  repeatability of the final spectral mapping products by analysing fully sampled observations of the Orion Bar, which were made every six months through the mission.

%
%
\begin{figure}
\centering
\includegraphics[width=0.46\textwidth]{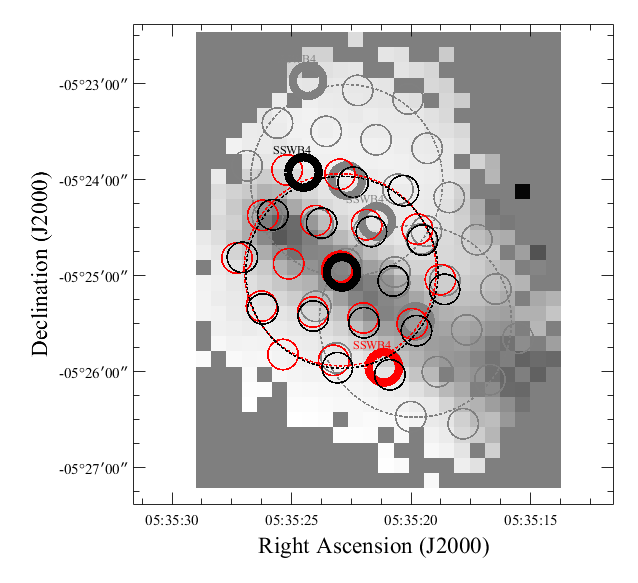}
\includegraphics[width=0.5\textwidth]{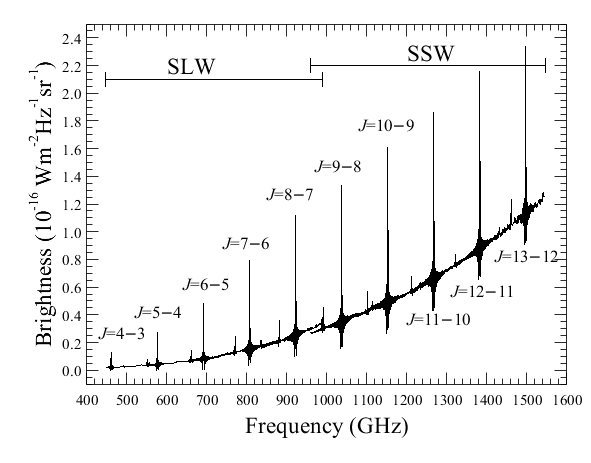}
\caption{The left plot shows thd" footprint of the unvignetted SPIRE SSW detectors from the centre jiggle position over the combined integrated spectral map. Four of the observations are included: OD302 with position angle 267 degrees (black); OD495 with position angle 89 degrees (red); and the two offset observations (grey). Two detectors, SSWD4 and SSWB4, are highlighted with thicker lines to show the rotation, and the 2$^{\prime}$ field of view is shown by the dashed circle. The right plot shows the average spectrum across the map from one of the observations. The strongest spectral lines, due to rotational transitions of $^{12}$CO, are labelled.}
\label{fig:orientation}       
\end{figure}

%
%
\section{Observations and data reduction}
\label{sec:2}

The Orion Bar was observed seven times during the {\it Herschel} mission, with observations separated in time by approximately six months. Each observation was centred on the star HD37041 at 5h 35m 22.9s, -5d 24$^{\prime}$ 57.8$^{{\prime}{\prime}}$ (J2000) and six were included in the SPIRE routine calibration program.  The seventh observation, from {\it Herschel} operational day (OD) 495, was made as part of the {\it Herschel} Key Program ``Evolution of Interstellar Dust'' \citep{abergel}. A preliminary analysis of a sparse FTS observation at the same pointing was previously presented by \citet{habart2010}. In addition, two further maps of the Orion Bar were made on OD302 with offsets of approximately 1$^{\prime}$ - i.e. with fields partially overlapping the seven co-pointed observations. All of the observations were carried out using the SPIRE FTS high spectral resolution, fully sampled mode, with two repetitions (each repetition is two scans) per jiggle position. As the Orion Bar is located close to the ecliptic, the position angle of the {\it Herschel} satellite during the observations falls into two groups, with angles of $\sim$270 degrees and $\sim$90 degrees. Table~\ref{tab:obs1} lists the observations with their offsets and position angles. The SPIRE SSW array footprint is shown on the sky in Fig.~\ref{fig:orientation} for four of the Orion Bar mapping observations, one from each group of observing angle and the two offset observations. Figure~\ref{fig:orientation} also shows a typical spectrum, with the main spectral features identified.

%
%
\begin{table}

\caption{Orion Bar observations. Offsets are relative to the position 5h 35m 22.9s, -5d 24$^{\prime}$ 57.8$^{{\prime}{\prime}}$. All observations were made with 2 repetitions per jiggle point.}
\label{tab:obs1}
\begin{tabular}{cccccc}
\hline\noalign{\smallskip}
Date & Operational & OBSID & Offset &  Position Angle  & Satellite Radial\\
  &  Day (OD) &       & (arcsec)  & (degrees)  & Velocity (km\,s$^{-1}$) \\
\noalign{\smallskip}\hline\noalign{\smallskip}
2010-03-11  &  302  & 1342192173 & 0.0, 0.0	& 266.50 & -44.8\\
2010-09-20  &  495  & 1342204919 & 0.0, 0.0     & 89.46  & 8.2\\
2011-02-27  &  655  & 1342214846 & 0.0, 0.0	& 260.33 & -44.0\\
2011-09-17  &  857  & 1342228734 & 0.0, 0.0	& 87.94  & 8.3\\
2012-03-10  &  1032 & 1342242591 & 0.0, 0.0	& 266.22 & -44.8\\
2012-08-12  &  1187 & 1342249465 & 0.0, 0.0	& 70.03  & 4.3\\
2013-03-04  &  1390 & 1342265845 & 0.0, 0.0     & 262.85 & -44.4\\
2010-03-11  &  302  & 1342192174 & -45.8, -30.6 & 266.52 & -44.8\\
2010-03-11  &  302  & 1342192175 & -3.0, 57.3   & 266.55 & -44.8\\
\noalign{\smallskip}\hline
\end{tabular}
\end{table}

%
%
\begin{figure}
\includegraphics[height=0.42\textwidth]{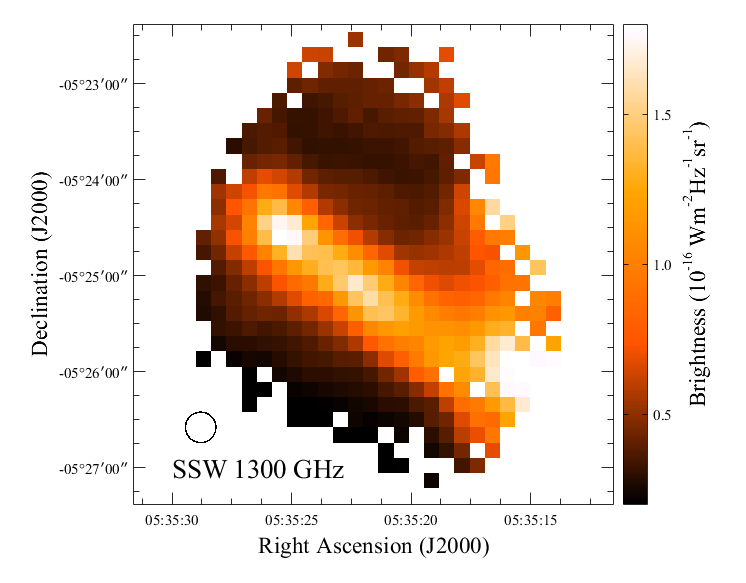}
\includegraphics[height=0.42\textwidth]{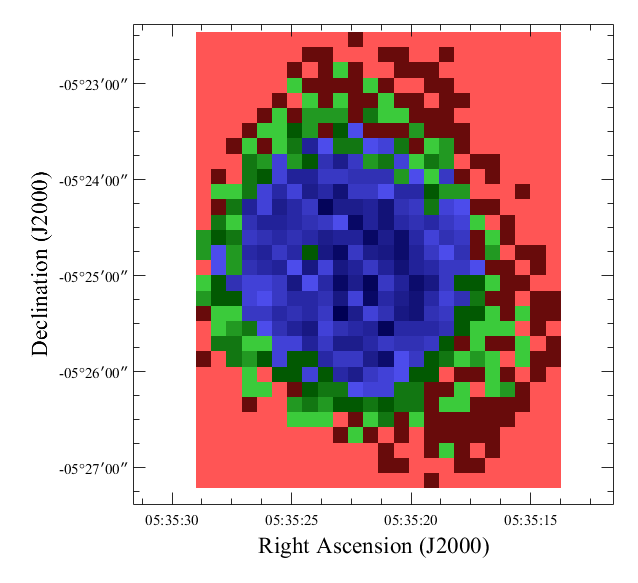}
\caption{The left plot shows the layer of the combined SSW cube at 1300~GHz. The approximate beam size is shown on the bottom left of the plot. The right plot shows the coverage map for the combined cube - bright red indicates zero coverage, dark red indicates pixels containing 4 scans, green indicates 8--20 scans, and blue 24--72 scans.}
\label{fig:orientation2}       
\end{figure}

The observations were processed using the standard {\it Herschel} Interactive Processing Environment \citep[HIPE; ][]{hipe} v11 pipeline, which applies a separate calibration to each jiggle position and detector. All positions were converted to units of W\,m$^{-2}$\,Hz$^{-1}$\,sr$^{-1}$ by the pipeline, which applies a calibration applicable for extended emission across the SPIRE beam \citep{swinyard13}. The assumption of extended emission used for the calibration can be tested by examining the spectrum in the overlap region between the SSW and SLW bands at $\sim$950~GHz. The beam FWHM for the two bands differs by approximately a factor of two, and so any structure in the source intensity distribution on this spatial scale will show up as a discontinuity. Figure~\ref{fig:orientation} shows the average spectrum over all pixels in one of the maps. The spectrum is approximately continuous between the two bands, indicating that the extended calibration is valid for this source.

In order to provide a reference against which to compare the individual maps, a combined (or ``mean'') spectral cube was constructed by gridding all of the observed sky points across the nine observations together. The spatial gridding was carried out using the standard SPIRE FTS pipeline task which uses a na\"{i}ve map making algorithm. The na\"{i}ve algorithm averages all of the detector samples that fall within each map pixel, and enters the corresponding number of FTS scans into the equivalent pixel in a coverage map. The grid spacing was set to 9.5$^{{\prime}{\prime}}$ for SSW and 17.5$^{{\prime}{\prime}}$ for SLW, which roughly represents half the beam size and matches the standard output of the data reduction pipeline. The image in the left plot of Figure~\ref{fig:orientation2} shows a layer of the combined cube for SSW at 1300~GHz; the plot on the right of Figure~\ref{fig:orientation2} shows the coverage for SSW. Each sky position in each observation was observed with 4 scans, but individual map pixels may also contain multiple pointings from a single observation, depending on how the rectangular grid falls over the hexagonal sky positions. The centre of the combined cube has the highest coverage, as all observations in this region overlap. 

An identical grid to the combined cube was used to produce individual spectral cubes for each observation, making it possible to calculate pixel-by-pixel ratios with the combined cube. The ratio cube for an observation $i$, $R_i(x,y,\nu)$, is defined as
\begin{equation}
R_i(x,y,\nu) = \frac{I_i(x,y,\nu)}{\overline{I(x,y,\nu)}},
\label{equ:ratio}    
\end{equation}
where $I_i(x,y,\nu)$ is the intensity cube of observation $i$, and $\overline{I(x,y,\nu)}$ is the mean intensity cube after combining all observations. 

An error cube is calculated by the na\"{i}ve algorithm as the standard error on the mean of samples within each map pixel. In order to be able to compare individual error cubes with the ratios derived above, the relative error cube is defined as
\begin{equation}
\sigma_{\mathrm{rel}}(x,y,\nu) = \frac{\sigma_i(x,y,\nu)}{I_i(x,y,\nu)},
\label{equ:noise}    
\end{equation}
where $\sigma_i(x,y,\nu)$ is the error cube from observation $i$.

%
%
\begin{figure}
\centering
\includegraphics[width=0.44\textwidth]{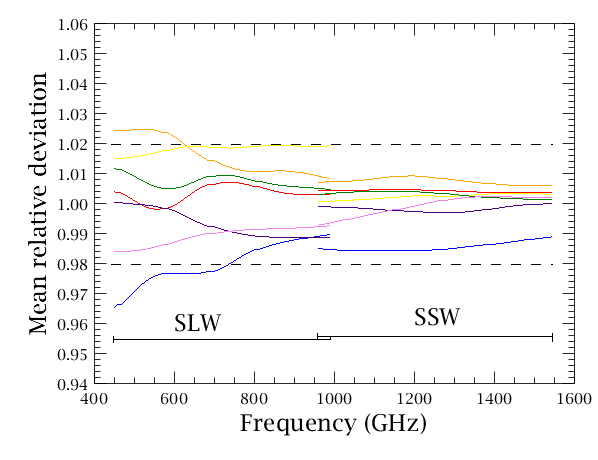}
\includegraphics[width=0.44\textwidth]{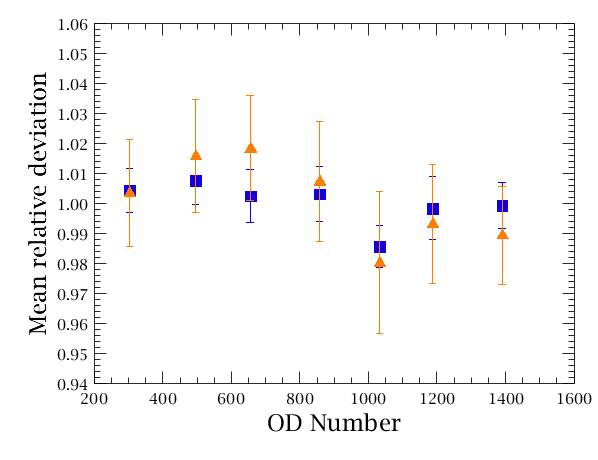}
\caption{The left plot shows the smoothed mean relative deviation spectrum for each observation (after masking the CO lines). The line colours correspond to different observations, identified by the {\it Herschel} operational day: OD302 (red), OD495 (orange), OD655 (yellow), OD857 (green), OD1032 (blue), OD1187 (purple), OD1390 (pink). The right plot shows the values averaged over all frequencies (avoiding spectral lines) plotted against Herschel observing day. The blue squares are for SSW and the orange triangles for SLW, and the error bars represent the standard error on the mean relative deviation. }
\label{fig:meanmean}
\end{figure}

%
%
\section{Relative calibration}
\label{sec:3} 

In order to investigate any systematic effects in the overall relative calibration of the seven co-pointed maps, and to determine the consistency of the maps over the course of the mission, the mean over all pixels in the ratio cube for observation $i$ was calculated as
\begin{equation}
\overline{r_i(\nu)} = \frac{\sum_{x,y}{R_i(x,y,\nu)}}{n_{x,y}},
\label{equ:mrd}    
\end{equation}
where $n_{x,y}$ is the number of pixels in the cube and $\overline{r_i(\nu)}$ is the mean relative deviation spectrum for each observation compared to the average of all observations. Only the seven co-pointed observations were used in order to compare the mean value of the ratio cube over the same area on the sky.

The resulting mean relative deviation spectrum for each observation is shown in the left panel of Fig.~\ref{fig:meanmean} after smoothing the frequency axis with a Gaussian filter of width (FWHM) 50~GHz. A $\sim$10~GHz region around each $^{12}$CO line was masked before applying the smoothing function. Masking is necessary in order to avoid large values in the ratio cube, assumed to be caused by small shifts in line position between observations (see Sect.~\ref{sec:6}).  Figure~\ref{fig:meanmean} also shows how the mean relative deviation changes with time through the mission. The error bars in this figure represent the standard error on the mean relative deviation.

The results show that the overall level of the observations is consistent, within $\pm$2\% and that there is no evidence of a systematic change in time through the mission. This shows that the mapping observing mode and relative calibration is stable.

%
%
\section{Repeatability}
\label{sec:4} 

\begin{figure}
\centering
\includegraphics[width=0.6\textwidth]{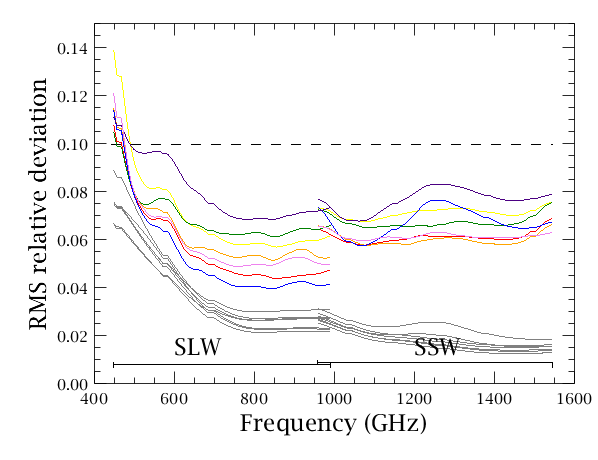}
\caption{The smoothed RMS deviation spectra for the seven co-aligned Orion observations (after masking the CO lines). The line colours correspond to the different observations - see Fig.~\ref{fig:meanmean}. The relative error spectrum is also shown (in grey) for each observation.}
\label{fig:rms}
\end{figure}

The variation in individual pixel values between each observation can be used to assess the repeatability of the spectral cubes. In order to summarise the results over many pixels, the spread of pixel values in the individual ratio cubes was examined using the root mean square (RMS) deviation from unity
\begin{equation}
\sigma_{\mathrm{rms}}(\nu) = \sqrt{\frac{\sum_{x,y}{(R_i(x,y,\nu)-1)^{2}}}{n_{x,y}}}.
\label{equ:rms}    
\end{equation}
This RMS deviation measures the spread in the pixel values in each observation compared to the mean over all observations. It incorporates: the intrinsic random uncertainty in making multiple measurements; the absolute pointing error (APE) of the {\it Herschel} telescope; the fact that the sky may have been sampled by different detectors (and jiggle positions) in different observations; and differences in the observed sky positions relative to the rectangular grid (and the source brightness distribution). The RMS deviation spectrum for each observation is plotted in Fig.~\ref{fig:rms} after masking the strong CO lines and smoothing as described earlier in Sect.~\ref{sec:3}. The values for SLW are in the range 4--7\% for frequencies above 600~GHz, but rise above 10\% towards the low frequency edge of the band where the emission from the source is lowest and uncertainty associated with the instrument background emission may be important (see Sect.~\ref{sec:5}). The value for SSW is $\sim$7\% across the band. In order to try to separate the contributions to this uncertainty, each effect listed above was investigated in more detail. 

%
%

The error cube, calculated by the na\"ive map making algorithm, contains the standard error on the mean of the individual samples that fall within each pixel. This error is due to the spread in repeated scans that contribute to any given pixel. The mean over all the pixels of the relative error cube can therefore be used as a measure of the random uncertainty plus the difference in sky positions with respect to the pixel grid. This relative error was calculated using Equation~\ref{equ:noise} and compared to the RMS deviation for each observation in Fig.~\ref{fig:rms}. The relative error has a value of $\sim$2\%, which is lower than the RMS deviation, confirming there are additional contributions to the RMS deviation, above the random spread of repeated scans.

The APE of the {\it Herschel} telescope is approximately 2$^{{\prime}{\prime}}$ at 68\% confidence \citep{miguel}. In order to test the effect of pointing on the RMS deviation, a relative pointing offset was determined from the SSW cubes (i.e. using the smallest beam) such that it minimised the residual between different observations. These offsets were all found to be consistent with the expected APE of the telescope with values in the range 2--4$^{{\prime}{\prime}}$. The shifts were then applied to the pointing of each individual detector/jiggle position spectrum, before gridding into the cube. As the calculated shifts were all small compared to the grid size, the RMS deviation for the pointing-aligned observations was not significantly different with or without the shifts applied, which shows the pointing uncertainty from the telescope APE is not a significant source of error contributing to the RMS deviation.

In order to check the detector-detector calibration, the individual spectra were examined before gridding, and clipped spectra, where the interferogram signal went out of the detector dynamic range, were classified as bad and removed. In addition to this, the standard pipeline restricts the map area by excluding partially vignetted detectors in the outer ring of each array \citep[see][ for a definition of the vignetted detectors]{griffin2010}. Figure~\ref{fig:rms} shows the RMS deviation after removing obviously bad spectra and restricting the cube to include only unvignetted detectors. 

The two offset observations, from OD302, also provide a useful check on the detector-detector calibration, as the central part of the source lies in different parts of the array. The RMS deviation can be calculated for these observations in the region where they overlap with the co-aligned maps. The value of the RMS increases to $\sim$9\% for SSW and SLW, indicating that the uncertainty in the map increases towards the edges of the array. Including the vignetted detectors has a similar effect, increasing the RMS deviation.

The final effect that could contribute to the RMS deviation is the differences in the observed sky positions relative to the grid. This is important for SPIRE Spectrometer data because there are generally only 1--3 observed positions on sky within each map pixel - i.e. the redundancy in the observed sky points is low. This means that if the source brightness distribution varies on the scale of the pixel size, there could be uncertainties introduced to the pixel value depending on exactly where the observed sky points were placed.

%
%
\section{Additional uncertainties}
\label{sec:5} 

An additional cause of uncertainty in mapping observations, not covered in the metrics and discussion in Sect.~\ref{sec:4}, is the accuracy with which the telescope and instrument emission is subtracted from the spectra \citep[in the same way as for sparse observations, see ][]{swinyard13}. This is difficult to investigate with the Orion Bar observations because the source is so bright. However, there can be residual additive contributions in the fainter parts of other mapping observations. Analysis of mapping observations towards a dark region of sky shows that the residual uncertainty is (0.2--0.7)$\times10^{-19}$ W\,m$^{-2}$\,Hz$^{-1}$\,sr$^{-1}$ for SLW and (0.7--1.7)$\times10^{-19}$ W\,m$^{-2}$\,Hz$^{-1}$\,sr$^{-1}$ for SSW \citep{swinyard13}.

%
%
\section{Frequency calibration}
\label{sec:6} 

\begin{figure}
\includegraphics[width=0.52\textwidth] {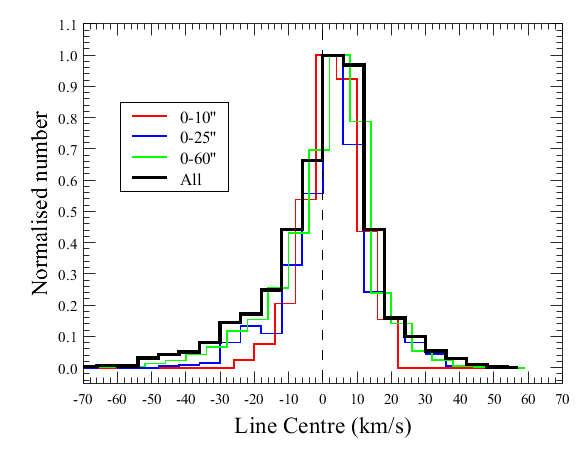}
\includegraphics[width=0.52\textwidth] {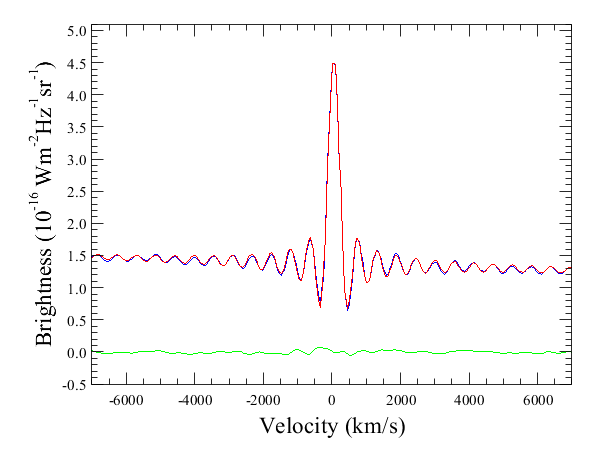}
\caption{ The left plot shows the normalised histogram of fitted $^{12}$CO line centres in the seven co-pointed spectral maps of Orion, after subtracting 10~km\,s$^{-1}$ for the velocity of the source. The solid black line shows the histogram for all pixels in the maps, and the other lines show the histogram for pixels at varying radii from the centre of the map. The right plot shows an example line fit for the $^{12}$CO $J$=11--10 with data in blue, sinc fit in red and residual in green.}
\label{fig:lines}
\end{figure}

The frequency scale of the SPIRE FTS is determined by the position of the moving FTS mirror (the SMEC) and the associated uncertainty is determined by how accurately the SMEC positions can be measured and converted into optical path difference (OPD). The conversion between the measured SMEC position and corresponding OPD varies for each SPIRE detector, depending on the radial offset of that detector from the instrument optical axis \citep{spencer}. This ``obliquity'' effect is important for high spectral resolution mapping observations where different detectors and jiggle positions can be combined by the gridding process  described above (see Sect.~\ref{sec:2}). In the FTS data processing pipeline, obliquity effects are taken into account by applying a scale factor when converting between the SMEC mechanical position and the OPD for each detector. The details of how this scale factor was derived are given by \citet{swinyard13}. In addition, the obliquity effect causes a radially varying spectral resolution (i.e. line width) across the detector array \citep{spencer}. However, in this paper, we concentrate only on the accuracy of the frequency calibration.

In order to test the accuracy of the frequency calibration, the ten $^{12}$CO lines shown in Fig.~\ref{fig:orientation} were fitted in each of the spectral cubes produced by the pipeline for the seven co-pointed observations of the Orion Bar. For each pixel in each spectral cube, the spectrum was fitted simultaneously with a second order polynomial for the continuum, and a sinc function for each of the $^{12}$CO lines \citep[the line shape for the SPIRE FTS is well fitted by a classical sinc function; e.g.][]{naylor10}. An example fit for the $^{12}$CO $J$=11--10 line is shown in the right hand side of Fig.~\ref{fig:lines}.  The lines are well separated and so the fit is not affected by blending of the line profile wings or by the weaker $^{13}$CO lines. It was necessary to apply a correction for the velocity of the satellite relative to the Local Standard of Rest (LSR), as this is not performed by the standard pipeline in HIPE v11. Table~\ref{tab:obs1} shows the satellite velocity for each observation.

 The intrinsic width of the CO lines in the Orion Bar is narrow \citep[$\sim$3~km\,s$^{-1}$, e.g.][]{wilson} compared to the instrumental spectral resolution of the FTS, which varies from 800~km\,s$^{-1}$ at the low frequency end of the band to 230~km\,s$^{-1}$ at the high frequency end. The spectral noise level in each map pixel was calculated as the root mean square spread of the fit residual in a frequency range avoiding the strong CO line positions. The resulting signal-to-noise ratios (SNR) of the measured lines have a mean value of $\sim$90 and vary up to $\sim$380 for the strongest lines. With SNR values of this order, an estimate of the accuracy expected on the fitted line centres can be obtained from the line width divided by the SNR, assuming that the instrumental line shape is well known \citep[e.g.][]{davis}. 

Figure~\ref{fig:lines} shows the distribution of fitted line centres observed in SSW and SLW for the seven observations after subtracting 10~km\,s$^{-1}$ for the intrinsic velocity of the Orion Bar \citep[e.g. ][]{buckle2012}. Only spectral lines with a SNR greater than 90 (equivalent to an expected accuracy of the fitted line centres of 3--9~km\,s$^{-1}$) were included. This limit effectively excludes the $J$=4--3 and $J$=5--4 lines, which have the lowest SNR. In order to show how the accuracy changes towards the map edges, the results are plotted for pixels contained within various radii from the centre of the map. It should be noted that distance from the centre of the map does not directly correspond to offset from the optical axis due to the jiggle pattern and the fact that more than one detector may have been averaged together in each map pixel. The standard deviation of each distribution is: 8~km\,s$^{-1}$ ($<$10$^{{\prime}{\prime}}$); 12~km\,s$^{-1}$ ($<$25$^{{\prime}{\prime}}$); 14~km\,s$^{-1}$ ($<$60$^{{\prime}{\prime}}$); 15~km\,s$^{-1}$ (for all pixels in the map). There is some indication that the standard deviation is slightly smaller than this at the low frequency end of each band and slightly larger towards the high frequency end. 

The real variation in line velocities in the region of the Orion Bar is smaller than the measured spread in line centres shown in Fig.~\ref{fig:lines}. In the clumps along the bar, the velocity has been measured to be between 9.5 and 11~km\,s$^{-1}$ \citep{lisSchilke}, and $^{12}$CO $J$=7-6 maps of the wider region show very little emission outside of the range 8--14~km\,s$^{-1}$ \citep{wilson}. This small intrinsic velocity range indicates that the increasing spread in line centre distribution with radius is due to the uncertainty on the frequency calibration increasing towards the edge of the map. This spread, and also the slight asymmetry in the line centre distribution, are probably due to inaccuracies in the SMEC scale factor away from the optical axis (although radius from the map centre does not directly correspond to distance from the optical axis, as mentioned above). 

The spread in line velocities has also been measured for the most point-like evolved stars in the SPIRE routine calibration programme, and was found to be $<$7~km\,s$^{-1}$ \citep{swinyard13}. These stars were observed on the centre pair of detectors, and the value is close to the spread of 8~km\,s$^{-1}$ found for the centre part of the map. The agreement between the map centre and the point-source observations is consistent with the increased uncertainty at the map edges being due to the other off-axis detectors that are included for mapping observations.

%
\section{Conclusion}
\label{sec:6}  

The uncertainties involved in SPIRE FTS spectral mapping observations have been investigated using pipeline data corresponding to HIPE v11 for the Orion Bar, which was observed through the mission at intervals of six months. The main conclusions can be summarised as follows:
\begin{itemize}
\item The mapping mode was stable through the mission with no trend in time between the different observations
\item The RMS deviation of each observation from the combination of all observations was 4--7\% for SLW (above 600~GHz) and 7\% for SSW
\item The uncertainty is probably due to detector-to-detector calibration, with the uncertainty rising towards the edge of the array, and particularly when the partially vignetted detectors in the outer ring are included
\item The frequency calibration of mapping observations has a larger uncertainty towards the edges of the map than for single pointed observations, due to combining the data from different detectors and jiggle positions. The uncertainty in line velocity for the whole map is $\sim$15~km\,s$^{-1}$.
\end{itemize}

In future versions of HIPE, we hope to improve the detector-detector calibration such that the partially vignetted detectors can be safely included in the maps, with bad data automatically removed. In addition, we hope to understand and make corrections for the spread in line frequencies.

\begin{acknowledgements}
\textit{Herschel} is an ESA space observatory with science instruments provided by European-led Principal Investigator consortia and with important participation from NASA. SPIRE has been developed by a consortium of institutes led by Cardiff University (UK) and including Univ. Lethbridge (Canada); NAOC (China); CEA, LAM (France); IFSI, Univ. Padua (Italy); IAC (Spain); Stockholm Observatory (Sweden); Imperial College London, RAL, UCL-MSSL, UKATC, Univ. Sussex (UK); and Caltech, JPL, NHSC, Univ. Colorado (USA). This development has been supported by national funding agencies: CSA (Canada); NAOC (China); CEA, CNES, CNRS (France); ASI (Italy); MCINN (Spain); SNSB (Sweden); STFC (UK); and NASA (USA).
\end{acknowledgements}



\end{document}